\documentclass[12pt,english,amsmath,amssymb,preprint]{revtex4}
\usepackage[T1]{fontenc}
\usepackage[latin9]{inputenc}
\usepackage{color}
\usepackage{babel}
\usepackage{graphicx}
\usepackage{esint}
\usepackage[unicode=true,pdfusetitle,
 bookmarks=true,bookmarksnumbered=false,bookmarksopen=false,
 breaklinks=false,pdfborder={0 0 1},colorlinks=false,pdfpagemode=None]
 {hyperref}
\hypersetup{
 bookmarks,colorlinks,citecolor=red,filecolor=blue,urlcolor=blue}

\makeatletter
\@ifundefined{textcolor}{}
{%
 \definecolor{BLACK}{gray}{0}
 \definecolor{WHITE}{gray}{1}
 \definecolor{RED}{rgb}{1,0,0}
 \definecolor{GREEN}{rgb}{0,1,0}
 \definecolor{BLUE}{rgb}{0,0,1}
 \definecolor{CYAN}{cmyk}{1,0,0,0}
 \definecolor{MAGENTA}{cmyk}{0,1,0,0}
 \definecolor{YELLOW}{cmyk}{0,0,1,0}
 }

\usepackage{epsfig}% Include Figure files
\usepackage{dcolumn}% Align table columns on decimal point
\usepackage{subfigure}\usepackage{ulem}\usepackage{bm}% bold math
\@ifundefined{definecolor}{\usepackage{color}}{}
\usepackage{latexsym}\usepackage{latexsym}%%%%%%%%%%%%%%%%%%%%%%%%%%%%%%%%%%%%%%%%%%%%%%%%%%%%%%%%%%%%%%%%

\makeatother

\begin{document}

\title{Nematic order in the vicinity of a vortex in superconducting FeSe}

\author{Debanjan Chowdhury}

\affiliation{Department of Physics, Harvard University, Cambridge MA 02138}

\author{Erez Berg}

\affiliation{Department of Physics, Harvard University, Cambridge MA 02138}

\author{Subir Sachdev}

\affiliation{Department of Physics, Harvard University, Cambridge MA 02138}
\begin{abstract}
We present a phenomenological theory of the interplay between nematic order and superconductivity in the vicinity of a vortex induced by an applied magnetic field. Nematic order can be strongly enhanced in the vortex core. As a result, the vortex cores become elliptical in shape. For
the case where there is weak bulk nematic order at zero magnetic field,
the field-induced %EB nematicity
eccentricity of the vortex core has a slow power-law decay away from
the core. %EB, and so can lead to strong nematic order across
%the entire vortex lattice.
Conversely, if the nematic order is field-induced, then the eccentricity
is confined to the vortex core. We discuss the relevance of our results
to recent scanning tunneling microscopy experiments on FeSe (Song
\textit{et al.}, Science \textbf{332}, 1410 (2011)).
\end{abstract}

\date{\today}

\maketitle
%%%%%%%%%%%%%%%%%%%%%%%%%%

\section{Introduction}

The unconventional superconductors have a rich phase diagram
determined by the interplay of multiple competing, or coexisting,
types of order. Nematic order (which breaks the C$_{4}$ symmetry
of the underlying square lattice down to C$_{2}$) has been shown
to emerge in certain regimes of the phase diagrams of the
copper-oxide
\cite{ando02,hinkov08a,kohsaka07,taill10b,lawler,mesaros} and the
iron-based \cite{pnic1,pnic2,pnic3,pnic4,ming,uchida,chen}
superconductors. In the latter case, the nematic order accompanies
(and in some cases, precedes) the magnetic order which occurs at a
wavevector that breaks the lattice rotational symmetry.

Recently, the structure of the vortex cores in the mixed state of
clean FeSe films was studied by means of scanning tunneling microscopy
(STM) \cite{song11}. Strong anisotropy was observed in the zero bias
conductance map around the cores, which have an eccentricity of the
order of unity. Although the lattice structure of FeSe at low temperature
is orthorhombic\cite{mcqueen}, it has been claimed \cite{song11}
that the crystalline anisotropy (of the order of a few tenths of a
percent) is too small to explain the large anisotropy of the vortex
cores, which is likely to have an electronic origin.

This experiment raises several questions, some of which we address
in this paper: assuming that there is an electronic nematic order
in superconducting FeSe, what is its microscopic origin? What is its
relation to superconductivity - \textit{e.g.\/}, are these two types
of order competing? Is the nematic order localized in the vortex cores
(and hence stabilized by the application of the magnetic field), or
does it extend throughout the system (and is only apparent in the
STM spectrum near the cores)?

Here, we study the structure of the vortex core using a phenomenological
Landau-Ginzburg (LG) theory in terms of two competing order parameters.
Using our LG analysis we have calculated the structure of an isolated
vortex in the presence of the nematic order. Our main result is that
by looking at the profile of the gap near the vortex core, it is possible
to distinguish between two different configurations of the nematic
order, namely the presence of a localized nematic order within the
superconducting vortex as opposed to the presence of a long range
nematic order in the system. If the nematic order is localized at
the core, the superconducting gap should be anisotropic only near
the core and the anisotropy decays exponentially as we move away from
the core. On the other hand, if the nematic order is long-ranged,
the superconducting gap should exhibit an anisotropy which decays
as a power law. If the nematic order is near its critical point, there
is a large region in which the anisotropy of the gap depends logarithmically
on the distance, eventually crossing over to a power law. Moreover,
we find qualitative differences in the shape of the contours of constant
gap around the core in the different cases. If the nematic order exists
only in the cores, the equal-gap contours tend to be elliptical; if
the nematic order is long-ranged, we find that the gap function tends
to develop a {}``four-lobe'' structure, with more pronounced higher
harmonics. These features can be sought in STM experiments by mapping
the magnitude of the gap around the core as a function of position.

The paper is organized as follows: In section \ref{mod} we
introduce the LG functional with the two competing order
parameters and carry out a preliminary analysis in the absence of
the anisotropy. In section \ref{pd}, we investigate the mean-field
phase diagram of a single vortex. In section \ref{min}, we
introduce the anisotropy
and perform a %EBfull
numerical minimization of the functional, commenting on the interesting
features. Finally, in section \ref{anis}, we present our analytical
results explaining the various interesting features observed by minimizing
the free energy.

\section{Model}

\label{mod} We consider a LG type free energy for two competing order
parameters: a complex field $\Psi$, describing the superconducting
order parameter, and a real field $\phi$, which describes a nematic
order that competes with the superconducting order parameter. The
form of the free energy density is given by
\begin{eqnarray}
{\cal {F}} & = & {\cal {F}}_{s}+{\cal {F}}_{\phi}+{\cal {F}}_{a}+\frac{\gamma}{2}|\Psi|^{2}\phi^{2},\\
{\cal {F}}_{s} & = & \frac{\kappa_{\psi}}{2}|(-i\nabla-e^{*}{\bf {A}})\Psi|^{2}-\frac{\psi_{0}^{2}}{2}|\Psi|^{2}+\frac{1}{4}|\Psi|^{4},\\
{\cal {F}}_{\phi} & = & \frac{\kappa_{\phi}}{2}(\nabla\phi)^{2}-\frac{\phi_{0}^{2}}{2}\phi^{2}+\frac{1}{4}\phi^{4},\\
{\cal {F}}_{a} & = & \frac{\lambda_{1}}{2}\phi\bigg[|(-i\partial_{x}-e^{*}A_{x})\Psi|^{2}-|(-i\partial_{y}-e^{*}A_{y})\Psi|^{2}\bigg]+\frac{\lambda_{2}}{2}\phi\bigg[(\partial_{x}\phi)^{2}-(\partial_{y}\phi)^{2}\bigg].\label{GLfun}
\end{eqnarray}
 Apart from the standard free energy contributions arising due to
$\phi$ and $\Psi$, we have a competition term, controlled by $\gamma$
($>0$), and a term that gives rise to different effective masses for $\Psi$
in the two directions, which is controlled by $\lambda_{1}$. %EBThere
%is an underlying symmetry in ${\cal {F}}_{a}$, given by
${\cal {F}}$ is invariant under a rotation by 90 degrees, represented
by
\begin{eqnarray}
x & \rightarrow & y,\nonumber \\
y & \rightarrow & -x,\nonumber \\
\phi & \rightarrow & -\phi.
\end{eqnarray}
 We will be interested in the limit of $\Lambda\to\infty$, where
$\Lambda$ is the London penetration depth, so that we can neglect
the coupling to the electromagnetic field. At the outset, we set
$\lambda_{2}=0$, since the $\lambda_2$ term is small compared
to the $\lambda_{1}$ term in the limit where $\phi$ is small. It
is convenient to define the coherence length of $\Psi$ and the
healing length of $\phi$ as
\begin{equation}
l_{\psi}=\sqrt{\frac{\kappa_{\psi}}{\psi_{0}^{2}}},l_{\phi}=\sqrt{\frac{\kappa_{\phi}}{\phi_{0}^{2}}}.
\end{equation}
 Taking the unit of distance to be $l_{\phi}$, we can recast the
above free energy in a more transparent form as follows,
\begin{eqnarray}
{\cal {F}} & = & \frac{1}{2l^{2}}(\tilde{\nabla}\tilde{\psi}^{*})(\tilde{\nabla}\tilde{\psi})-\frac{1}{2}|\tilde{\psi}|^{2}+\frac{1}{4}|\tilde{\psi}|^{4}\nonumber \\
 & + & \bigg(\frac{\gamma}{\gamma_{s}}\bigg)^{2}\bigg[\frac{1}{2}(\tilde{\nabla}\tilde{\phi})^{2}-\frac{1}{2}\tilde{\phi}^{2}+\frac{1}{4}\tilde{\phi}^{4}\bigg]+\frac{\gamma^{2}}{2\gamma_{s}}|\tilde{\psi}|^{2}\tilde{\phi}^{2}\nonumber \\
 & + & \lambda\tilde{\phi}[(\partial_{\tilde{x}}\tilde{\psi}^{*})(\partial_{\tilde{x}}\tilde{\psi})-(\partial_{\tilde{y}}\tilde{\psi}^{*})(\partial_{\tilde{y}}\tilde{\psi})],\label{LGF}
\end{eqnarray}
 where $l=l_{\phi}/l_{\psi},\gamma_{s}=\gamma\psi_{0}^{2}/\phi_{0}^{2},\lambda=\lambda_{1}/2l_{\phi}^{2}\psi_{0}^{2}$,
$\tilde{x},\tilde{y}=x/l_{\phi},y/l_{\phi}$, $\tilde{\psi}=\Psi/\psi_{0}$,
and $\tilde{\phi}=\phi/\phi_{0}$. From now on, we will drop the tilde
symbols.

%EB
For $\lambda\ne0$, a short-distance cutoff has to be imposed on
Eq. \ref{LGF}. Otherwise, the system is unstable towards
developing modulations of $\psi$ with sufficiently short
wavelength. We discuss the instability in Appendix \ref{appins}.
In practice, we will mostly ignore this issue, assuming that there
is a short-distance cutoff (which is provided by the finite grid
used in our numerical calculations).

Before we begin our analysis, let us comment about the choice of parametrization
in this problem. We would like to think of this problem in terms of
a fixed $\gamma\leq1$. %EB: I think gamma<1 is necessary for the picture to hold. For gamma>1, if I got it right, there is a first order transition from a superconductor to nematic.
Then on choosing a particular ratio of the length scales of $\phi$
and $\psi$, we still have one degree of freedom left in terms of
the masses or the stiffnesses of the two order parameters, which is
fixed by tuning $\gamma_{s}$.

If we assume that $\gamma_{s}>1$, %EBand $\gamma_{s}>\gamma$,
then the uniform ground state is given by $\psi=1$ and $\phi=0$.
This also constrains $\phi$ to be localized around the vortex cores,
by making the mass term for $\phi$ positive deep inside the superconducting
region. If we further assume that the nematic order is small, such
that $\frac{\gamma}{2}\phi^{2}\ll\psi_{0}^{2}$, then we can essentially
ignore the feedback of $\phi$ on $\psi$. Therefore, we will first
find the full profile of $\psi=\Psi_{0}$, the isolated vortex solution,
in the absence of the nematic order and use that to find the form
of the nematic order. Then $\Psi_{0}$ satisfies the following asymptotic
relations:

\begin{eqnarray}
\Psi_{0}(\rho) & \approx & \bigg[1-\frac{1}{2}\bigg(\frac{1}{l\rho}\bigg)^{2}\bigg]e^{i\theta},~~~~~~~~~~\rho\gg l^{-1}\\
\Psi_{0}(\rho) & \sim & Cl\rho e^{i\theta},~~~~~~~~~~~~~~~~~~~~~~~~~\rho\ll l^{-1}\label{asymp}
\label{psi0as}
\end{eqnarray}
 where $\rho=r/l_{\phi}$, $r$ being the radius in the original coordinate
system, and $C$ is a dimensionless constant. In general, it is difficult
to find the solution of the full LG equation for $\Psi_{0}$ for all
$\rho$ analytically. Therefore we obtain the vortex solution $\Psi_{0}=f(l\rho)e^{i\theta}$
for all $\rho$ by minimizing the functional in Eqn. {\ref{LGF}}
numerically in the absence of $\phi$.

The numerical solution conforms to the two asymptotic expressions
above. It is interesting to note that $\Psi_{0}$ does not recover
from the vortex core to its bulk value exponentially, but rather as
a power law \cite{Kivelson02,Sachdev01}. The behavior of $\phi(\rho)$
in the vicinity of a vortex with $\lambda=0$ was studied by Ref.~\onlinecite{Kivelson02}.

\section{Phase diagram}

\label{pd} 
\begin{figure}
\begin{centering}
\includegraphics[width=1\columnwidth]{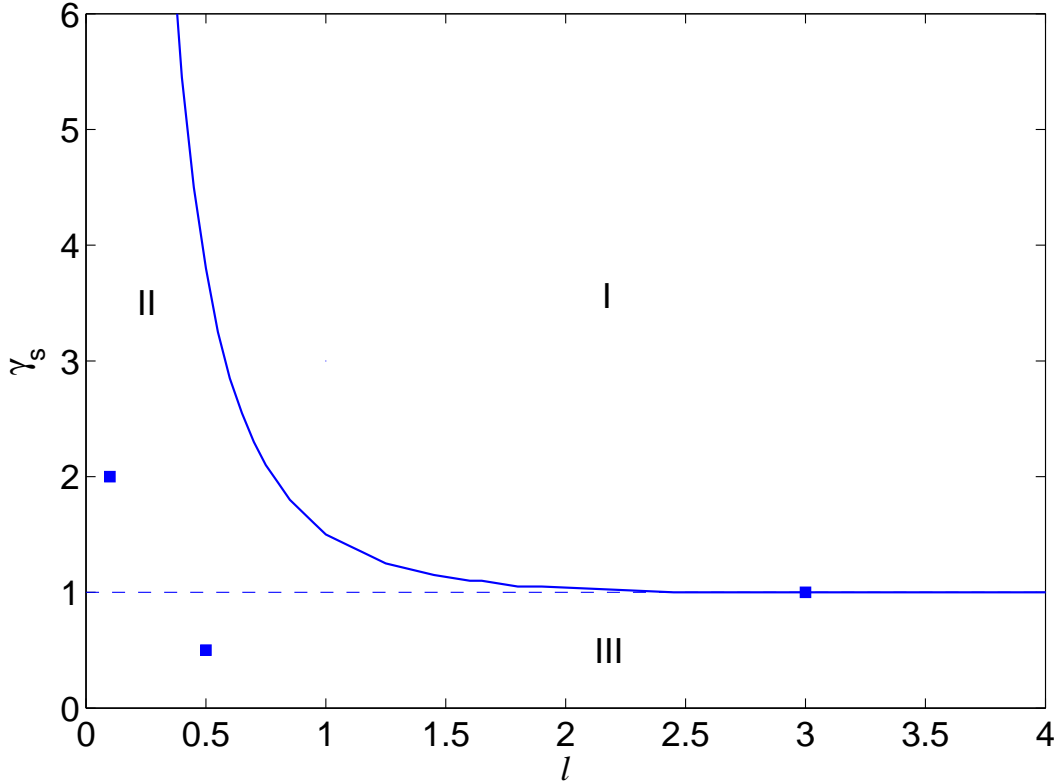}
\par\end{centering}

\caption{The phase diagram in the $(\gamma_{s},l)$ plane obtained by solving Eqn. \ref{eigeqn} numerically  on a grid with $n=100$
and $\Delta x=1$. The regions with qualitatively different solutions
for $\phi$ are marked. Phase I has no nematic order with $\phi=0$
everywhere, phase II has nematic order localized in the vortex core,
and phase III has long-range nematic order away from the core. The blue squares correspond to points which we
explore in more detail later. The dashed line $\gamma_{s}=1$ represents
the boundary between phases II and III.}

\label{lvsg}
\end{figure}

We will now describe the mean-field phase diagram of a
single vortex in the presence of a competing nematic order. There
are three possible phases: in phase I, $\phi=0$ everywhere; in phase
II, $\phi$ vanishes at large distance from the vortex core but becomes
non-zero near the vortex core due to the suppression of the competing
$\psi$ field to zero at the core; and in phase III, $\phi\ne0$ even
far away from the core. %EB Let us now address a very fundamental question, namely when is it possible to have a non-trivial % solution for  in the presence of the vortex solution.
%EB We would like to understand this in terms of a phase diagram in the
%$(\gamma_{s},l)$ plane.
A non-zero solution for $\phi$ is favored whenever the smallest eigenvalue
$\epsilon$ of the following eigenvalue problem \cite{Kivelson02}:
\begin{equation}
\bigg[-\nabla_{\rho}^{2}-1+\gamma_{s}[f(l\rho)]^{2}\bigg]\phi\left(x\right)=\epsilon\phi\left(x\right),
\label{eigeqn}
\end{equation}
 satisfies $\epsilon<0$. In order to find the phase diagram, we solve
this eigenvalue problem numerically on a discrete grid. The boundary
between phases I and II is the locus of points at which the smallest
eigenvalue satisfies $\epsilon=0$. For $\gamma_{s}<1$, $\phi$ becomes
long-ranged, corresponding to phase III. The resulting phase diagram
is shown in Fig. \ref{lvsg}. This phase
diagram is strictly valid as long as $\gamma_{s}>\gamma$. If this
is not the case, then the state with uniform nematic background and
no superconductivity is energetically favorable over any other state.

%%%%%%%%%%%%%%%%%%%%%%%%%%%%%%%%%%%%%%%%%%%%%%%%%%%%%%%%%%%%

%%%%%%%%%%%%%%%%%%%%%%%%%%%%%%%%%%%%%%%%%%%%%%%%%%%%%%%%%%%%

%EB Therefore, we can split the entire phase diagram into 3 distinct regions (see Fig. ): Region I %corresponds to where we only have the trivial solution . Region II corresponds to having a %localized solution for  inside the vortex core. Region III corresponds to having a background %nematic order coexisting with superconductivity. Finally the line  is where we have a quasi long %ranged solution for . An interesting question that we shall address later on is whether there is %a way to distinguish between the features of region II and region III using an experimental %probe. At this point, we would also like to stress that the whole picture of the above phase %diagram is strictly valid as long as . If this is not the case, then the state with uniform %nematic background and no superconductivity is energetically favorable over any other state.
% Let us now try and understand the features of this curve qualitatively. In the region , %non-trivial solutions exist only if . This includes solutions both from the region III and the %line , namely  being critical when  and there being a uniform background of  throughout the %system when .\\
The physics behind the phase diagram can be understood as follows.
When $l_{\phi}\gg l_{\psi}$ we are forcing the nematic order to
coexist with superconductivity in a large region. This is
unfavorable energetically due to the competition term
$\gamma\phi^{2}|\Psi|^{2}$. Therefore, when $\gamma_{s}>1$, there
is no $\phi\ne0$ solution. If $\gamma_{s}<1$, $\phi$ becomes
non-zero even far away from the vortex core. In the opposite limit
of $l_{\phi}\ll l_{\psi}$, the nematic order exists deep within
the superconducting vortex. Since there is very little overlap
between the two order parameters, the system can afford to have a
higher value of critical $\gamma_{s}$ below which there is a
nontrivial nematic order. This explains the increasing trend of
the critical
$\gamma_{s}$ for decreasing $l$. \\

In the $l_{\phi}\ll l_{\psi}$ case, it is possible to give an analytical
expression for the phase boundary between regions I and II. %EB In a similar fashion, we are also interested in obtaining the
%relationship between $\gamma_{s}$ and $l$ for small $l$.
%EB This can be obtained in a straightforward fashion using
 The equation for this curve is given by,
\begin{equation}
\gamma_{s}=\frac{1}{4(Cl)^{2}},
\end{equation}
where $C$ is the constant which appears in Eqn. \ref{psi0as}. The details of this computation are diskussed in Section \ref{phases}.

We are now in a position to include the effect of the anisotropy and
investigate the structure of the vortex cores in different regions
of the phase diagram described above.

\section{Vortex profile in the different regimes}

\label{min} We now turn to diskuss the characteristics of the vortex
profile in the different regimes shown in Fig. \ref{lvsg}. To solve
for the vortex profile, we minimize the free energy (\ref{LGF}) with
respect to $\psi$ and $\phi$ numerically on a disk geometry. This
is equivalent to solving the coupled Landau-Ginzburg equations with
Neumann boundary conditions, as we diskuss in Appendix \ref{boundary}.
Many of the features found in the numerical solution can be understood
analytically, as we diskuss in the next section.

We can expand both $\psi$ and $\phi$ in terms of the different angular
momentum channels ($\sim e^{in\theta}$). The term proportional to
$\lambda$ only couples angular momentum channels that differ by 2
units of angular momentum in $\psi$. %EB This can be
%seen from the fact the term $\lambda\phi(\partial_{x}^{2}-\partial_{y}^{2})$
%in plane polar coordinates gives rise to two terms which are proportional
%to $\lambda\phi\cos(2\theta)$ and $\lambda\phi\sin(2\theta)$.
Therefore, in the presence of $\phi$, the bare vortex solution ($\sim e^{i\theta}$)
gives rise to components of the form $e^{3i\theta}$, $e^{-i\theta}$,
etc. Similarly, the feedback of the superconducting order on $\phi$
gives rise to the generation of the even harmonics, i.e. $\Phi_{0}$
gives rise to terms proportional to $e^{2i\theta}$ and $e^{-2i\theta}$.
It is also possible to have a solution with only the even harmonics
of $\psi$, in which case, the vortex is absent. %EBIt is important to
%note that
These two solutions do not mix with each other and therefore we shall
focus on the solution in the presence of the vortex.\\

In light of this, we expand the order parameters as
\begin{equation}
\psi(\rho,\theta)=\sum_{n}\Psi_{n}(\rho)e^{i(2n+1)\theta},~~~~\phi(\rho,\theta)=\sum_{n}\Phi_{n}(\rho)e^{i2n\theta},~~~~n\in\mathrm{integers}\label{expan}
\end{equation}
In terms of the expansions in Eqn. \ref{expan}, the free energy
density can be written as,
\begin{eqnarray}
F_{\rho} & = & \int d\theta{\cal {F}}=\sum_{n}\frac{1}{2l^{2}}\bigg[\bigg(\frac{\partial\Psi_{n}}{\partial\rho}\bigg)^{2}+\frac{(2n+1)^{2}\Psi_{n}^{2}}{\rho^{2}}\bigg]-\frac{\Psi_{n}^{2}}{2}+\frac{1}{4}\sum_{n,p,q}\Psi_{n}\Psi_{n+p-q}\Psi_{p}\Psi_{q}\nonumber \\
 & + & \bigg(\frac{\gamma}{\gamma_{s}}\bigg)^{2}\bigg[\frac{1}{2}\bigg(\sum_{n}\bigg(\frac{\partial\Phi_{n}}{\partial\rho}\bigg)^{2}+\frac{(2n)^{2}}{\rho^{2}}\Phi_{n}^{2}-\Phi_{n}^{2}\bigg)+\frac{1}{4}\sum_{n,p,q}\Phi_{n}\Phi_{n+p-q}\Phi_{p}\Phi_{q}\bigg]\nonumber \\
 & + & \frac{\lambda}{2}\sum_{m,p}\phi_{p}\bigg[\bigg(\frac{\partial\Psi_{m}}{\partial\rho}-(2m+1)\frac{\Psi_{m}}{\rho}\bigg)\bigg(\frac{\partial\Psi_{m+p+1}}{\partial\rho}+(2(m+p+1)+1)\frac{\Psi_{m+p+1}}{\rho}\bigg)\nonumber \\
 & + & \bigg(\frac{\partial\Psi_{m}}{\partial\rho}+(2m+1)\frac{\Psi_{m}}{\rho}\bigg)\bigg(\frac{\partial\Psi_{m+p-1}}{\partial\rho}-(2(m+p-1)+1)\frac{\Psi_{m+p-1}}{\rho}\bigg)\bigg]\nonumber \\
 & + & \frac{\gamma^{2}}{2\gamma_{s}}\sum_{n,p,q}\Psi_{n}\Psi_{n+p-q}\Phi_{p}\Phi_{q},\label{LGhar}
\end{eqnarray}
 and we are interested in minimizing $\int\rho d\rho F_{\rho}$. We
shall minimize the above free energy for a given system size and
for only a fixed number of harmonics at a time. We have kept $n$
harmonics for $\psi$ and $\phi$, where for any given $n$ (odd) we
take all the harmonics $\Psi_{-i}$ to $\Psi_{i}$, $i=(n-1)/2$, and
similarly for $\phi$. We have tried $n=3,5$ and found no
substantial qualitative change in the results that we shall quote
here, indicating that the results converge even with only 3
harmonics. We consider a system on a disk of radius $\rho=100$.

Below, we describe the results in regions II and III of the phase
diagram, and on the critical line dividing them (In region I,
where there is no nematic order, we get the regular circularly
symmetric vortex). The specific values of $\gamma_{s},l$ which
were used are marked by blue squares in the phase diagram in Fig. {\ref{lvsg}}.

\subsection{Region II}

\begin{figure}
\psfig{figure=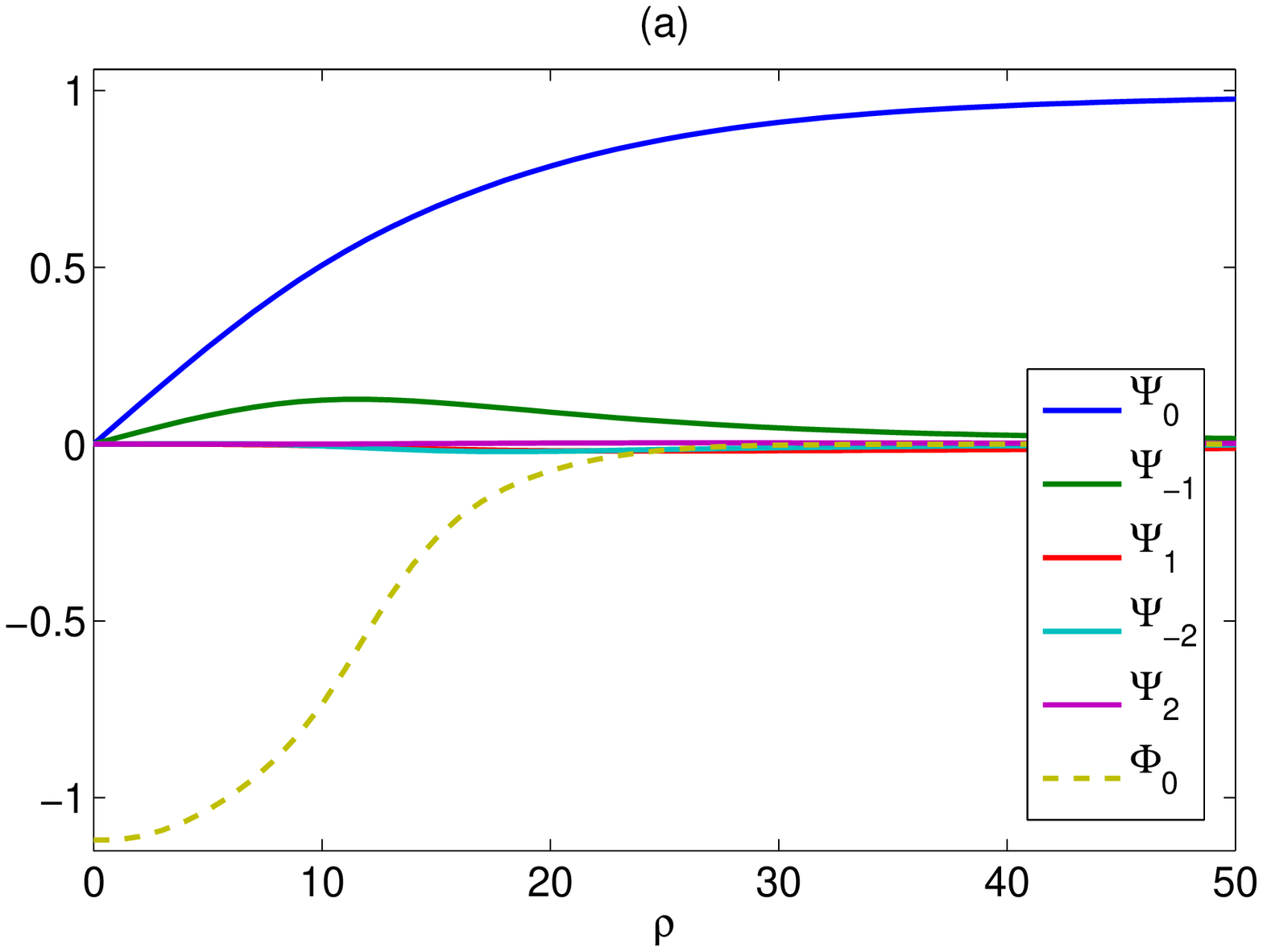,width=95mm} %\vspace{1mm}
\psfig{figure=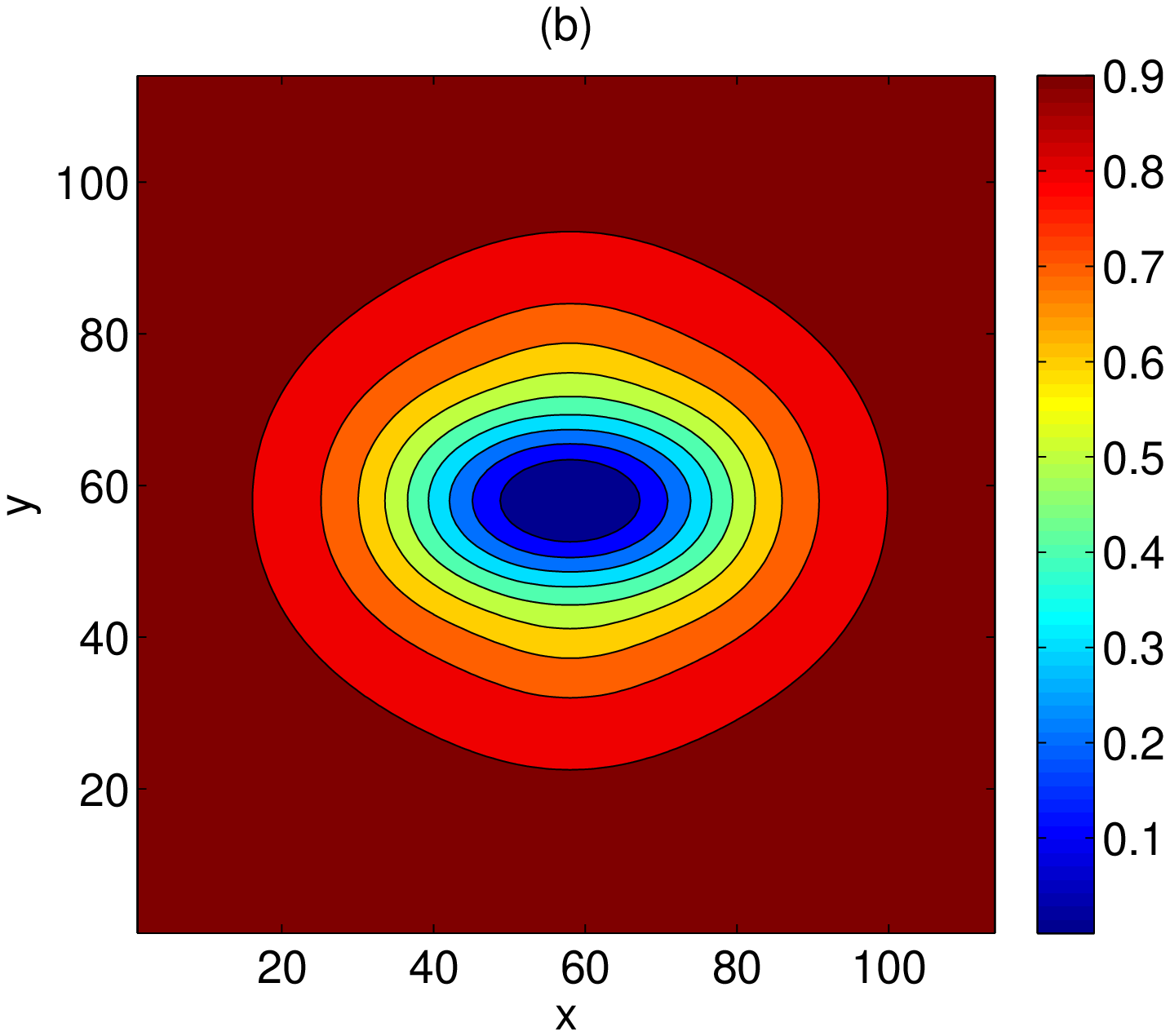,width=95mm} \vspace*{1mm}
 \caption{Nematic order in phase II. The profiles for the different order parameters
for $\gamma_{s}=2.0,l=0.1,\gamma=1.0,\lambda=20.0$ for a system size
of 100 ($N=100,\Delta x=1$). (a) Harmonics of $\psi$ (solid) and
$\Phi_{0}$ (dashed) (b) Contour plot of $|\psi|^{2}$. %EB The size of
%the vortex core in the absence of $\phi$
The superconducting coherence length $l_{\psi}$ is 10, in units of
$l_{\phi}$. }

\label{gam2l0p1}
\end{figure}

In this region, we expect to obtain a solution with a non-zero uniform
$|\psi|$ away from the vortex core and a non-zero $\phi$ localized
near the vortex core, decaying exponentially away from the core (given
that $\gamma<\gamma_s$ and $\gamma_{s}>1$). %EBat least for certain values of $l$ and $\gamma_{s}$. The profiles of $\phi$ and $\psi$ (decomposed into different harmonics)
The contour plot for $|\psi|^{2}$ is shown in Fig. {\ref{gam2l0p1}}.
The parameters used here are $\gamma_{s}=2.0,l=0.1,\gamma=1.0,\lambda=20.0$.
%\textbf{EB: I'm confused about the parameters here. I thought that for $\gamma>\gamma_{s}>1$, we get a solution with $\phi\ne0$ and $\Psi=0$.} {\color{red}{ Sorry, but this is a typo. $\gamma=1$ (See Fig. 3)}}

As can be seen in the figure, the core has an elliptical shape because
of the interaction with the nematic order which coexists with superconductivity
in the core region. As we go away from the core, the contours of equal
$|\psi|^{2}$ become more and more isotropic, due to the rapid decay
of the nematic order away from the core.

%%%%%%%%%%%%%%%%%%%%%%%%%%%%%%%%%%%%%%%%%%%%%%%%%%%%%%%%%%

%%%%%%%%%%%%%%%%%%%%%%%%%%%%%%%%%%%%%%%%%%%%%%%%%%%%%%%%%%%%%

\subsection{Region III}

\begin{figure}
\psfig{figure=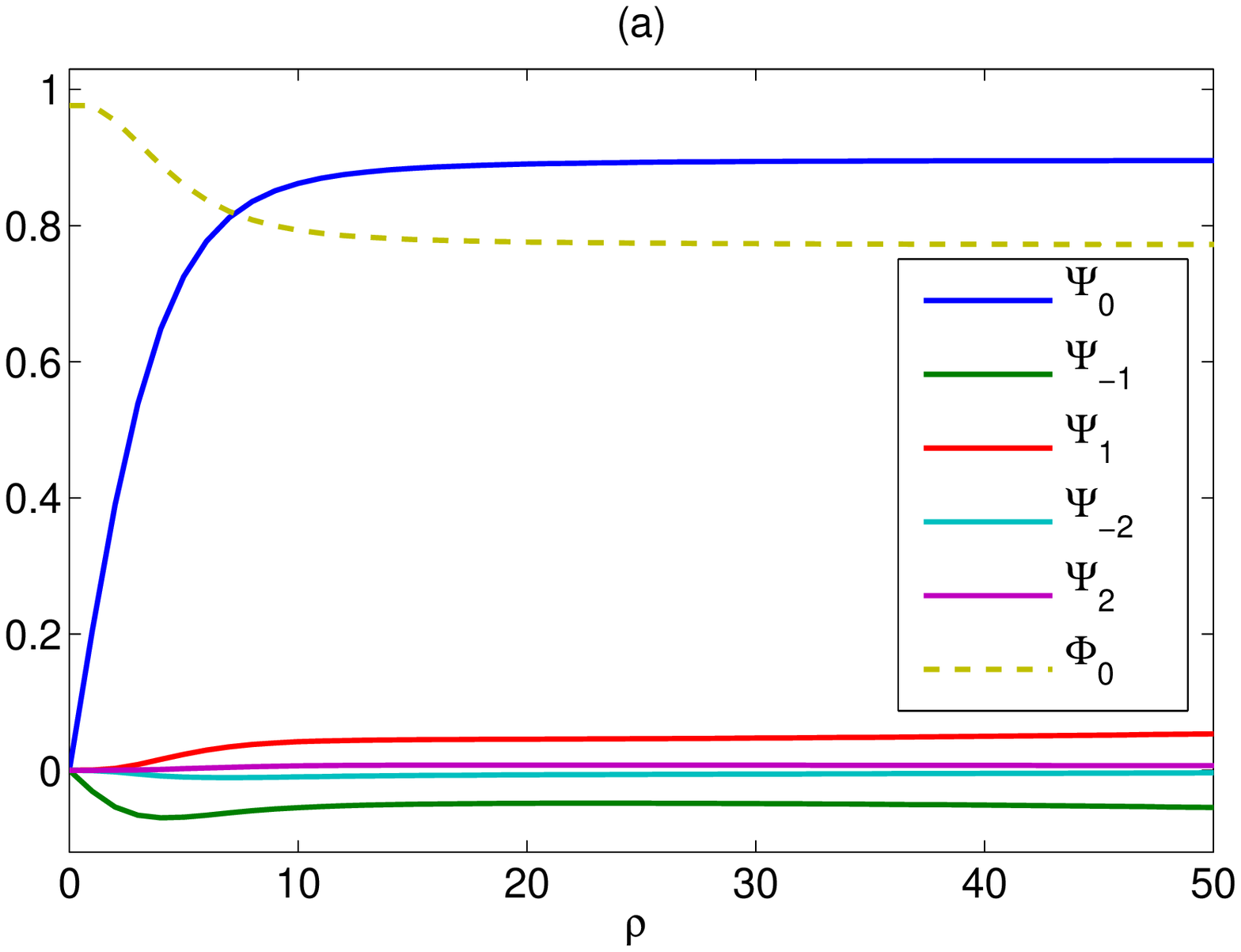,width=95mm} %\vspace{1mm}
\psfig{figure=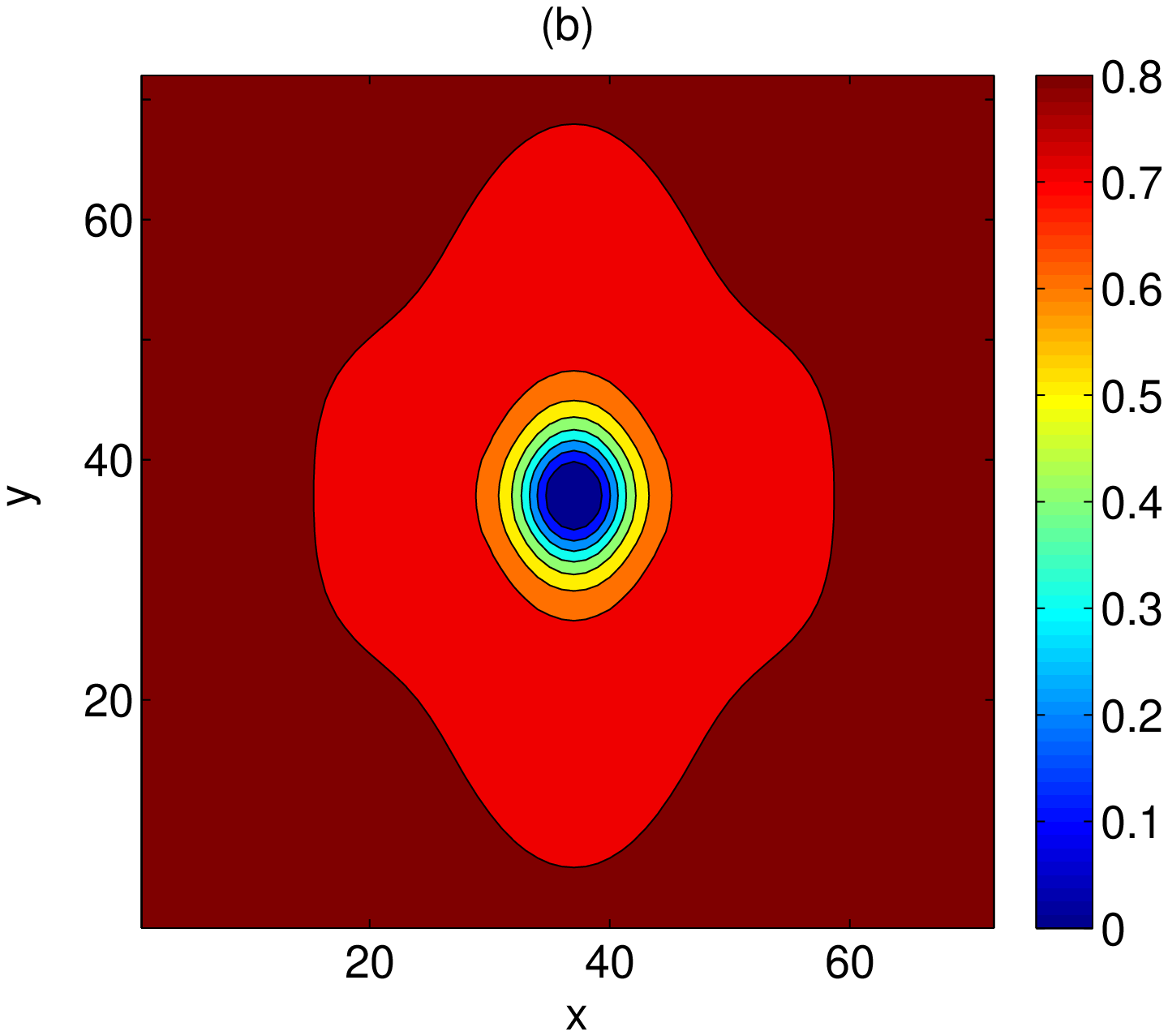,width=95mm} \vspace*{1mm}
 \caption{Nematic order in phase III. The profiles for the different order parameters
for $\gamma_{s}=0.5,l=0.5,\gamma=0.4,\lambda=0.5$ for a system size
of 100 ($N=100,\Delta x=1$). (a) Harmonics of $\psi$ (solid) and
$\Phi_{0}$ (dashed) (b) Contour plot of $|\psi|^{2}$. The superconducting coherence length is 2, in units of $l_{\phi}$. }

\label{gam0p5l0p5}
\end{figure}

This region in the phase diagram corresponds to the case where there
is a uniform nematic background coexisting with superconductivity,
even away from the vortex core. %EB, provided we
In this regime, as we move away from the core, $\phi$ goes to a constant
and $\psi$ remains anisotropic. In Fig. {\ref{gam0p5l0p5}}, the
harmonics $\Psi_{1}$ and $\Psi_{-1}$ are almost constant for large
$\rho$. Moreover, $\Psi_{1}=-\Psi_{-1}$ for large $\rho$. The contour
plot of $|\psi|^{2}$ reveals a large anisotropic {}``halo'' around
the core, with a non-elliptical shape.

Far away from the core, where $\phi$ is constant, the Landau-Ginzburg
equations can be solved analytically, showing that the anisotropy
in $|\psi|^{2}$ decays as a power law in this
case. We diskuss this solution in the next section.

%%%%%%%%%%%%%%%%%%%%%%%%%%%%%%%%%%%%%%%%%%%%%%%%%%%%%%%%%%

%%%%%%%%%%%%%%%%%%%%%%%%%%%%%%%%%%%%%%%%%%%%%%%%%%%%%%%%%%

\subsection{Critical case}

\begin{figure}
\psfig{figure=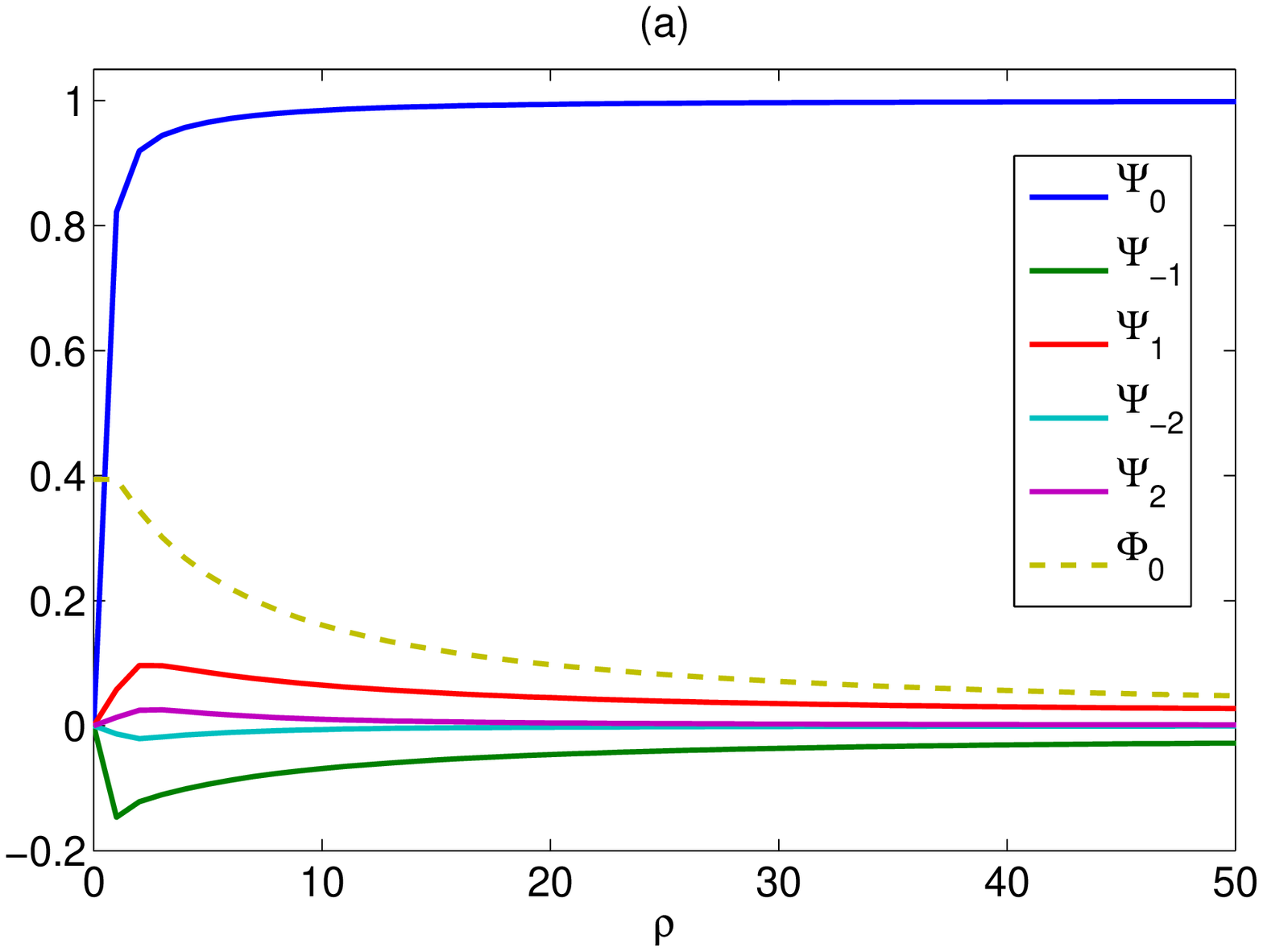,width=95mm} %\vspace{1mm}
\psfig{figure=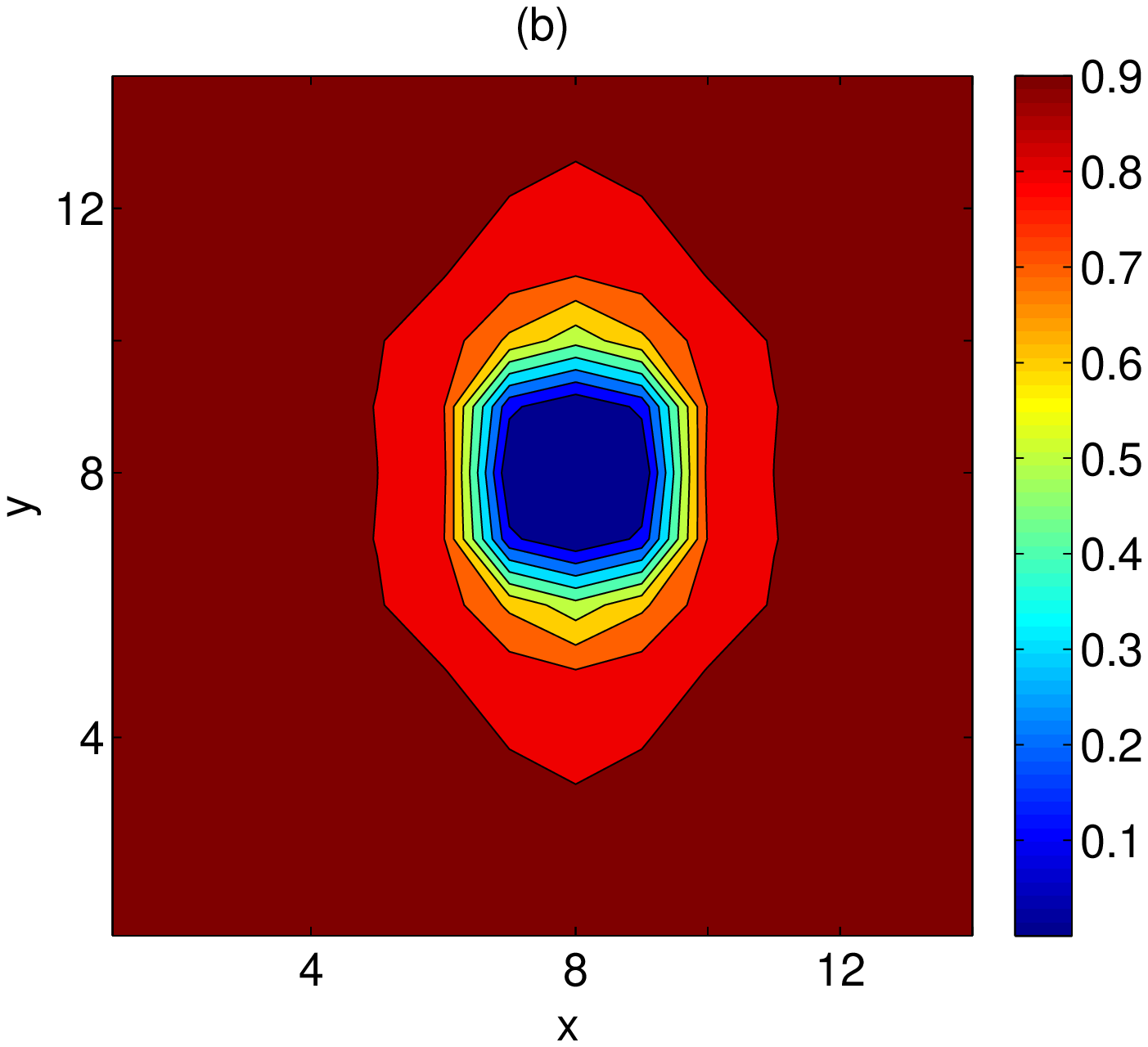,width=95mm} \vspace*{1mm}
 \caption{Nematic order at the critical point between phase III and phase I.
The profiles for the different order parameters for $\gamma_{s}=1.0,l=3.0,\gamma=0.9,\lambda=0.07$
for a system size of 100 ($N=100,\Delta x=1$). (a) Harmonics of $\psi$
(solid) and $\Phi_{0}$ (dashed) (b) Contour plot of $|\psi|^{2}$.
The superconducting coherence length is $1/3$, in units of $l_{\phi}$. }

\label{gam1l3}
\end{figure}

Finally, we diskuss the critical line separating regions II and
III in Fig. \ref{lvsg}, in which the $\phi$ field is critical far
away from the core. Naively, one would expect $\phi$ to go as
$1/\rho$ asymptotically in this regime. However, depending on
the details of the solution at small $\rho$, there may be
an intermediate regime in which $\phi\sim\ln\rho$, eventually
crossing over to $1/\rho$ at a larger distance. This feature is
diskussed in more detail in the next section. Fig. {\ref{gam1l3}}a
shows $\psi$ and $\phi$ in the
critical regime, with %\textbf{put parameters here}
$\gamma_{s}=1.0,l=3.0,\gamma=0.9$ and $\lambda=0.07$. Indeed,
we observe that $\phi$ decays slowly away from the core. $\Psi_{\pm1}$
also have long tails. The contour plot for $|\psi|^{2}$ shares features
that are similar to the behavior in region III, namely a long-range,
non-elliptical anisotropic halo. It is shown in Fig. {\ref{gam1l3}}b.

In the next section, we analyze the asymptotic behavior of the solution
in the critical case, showing that the anisotropic component of $|\psi|^{2}$
falls off as $\sim(\lambda\ln\rho/\rho^{2})\cos(2\theta)$ at intermediate
$\rho$, crossing over to $\sim(\lambda/\rho^{3})\cos(2\theta)$ at
sufficiently large $\rho$.

%%%%%%%%%%%%%%%%%%%%%%%%%%%%%%%%%%%%%%%%%%%%%%%%%%%%%%%%%%

%%%%%%%%%%%%%%%%%%%%%%%%%%%%%%%%%%%%%%%%%%%%%%%%%%%%%%%%%%%%%
%{\color{red}

\section{Analytical Treatment}

\label{anis} In this section, we propose various analytical
arguments to explain the different features that were observed
above by carrying out the minimization numerically. In subsection
\ref{phases}, we diskuss the solution for the nematic order in the
presence of superconductivity. Then in subsection \ref{coex}, we
analyze region III of the phase diagram (Fig. \ref{lvsg}). Finally,
in subsection \ref{lingl}, we study the linearized GL equations in
order to explain some of the other interesting features that were
observed earlier.

\subsection{Phases of the nematic order}

\label{phases} 
\begin{figure}
\begin{centering}
\includegraphics[width=0.6\columnwidth]{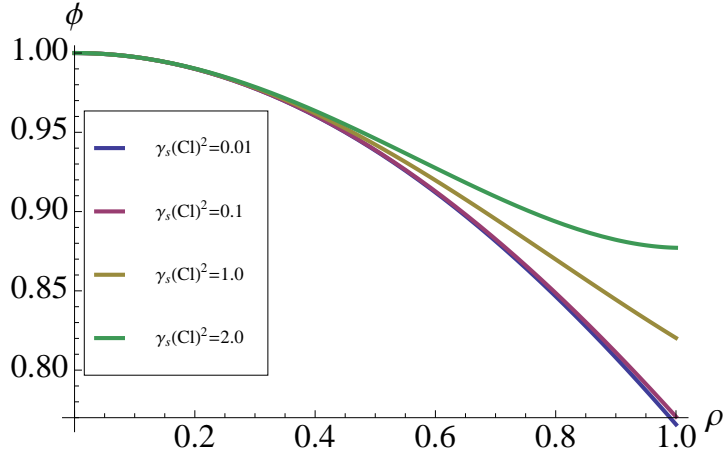}
\par\end{centering}

\caption{The profiles of $\phi$ as a function of $\rho$ for different values
of $\gamma_{s}C^{2}l^{2}$ over a distance of one correlation length
of $\phi$, $l_{\phi}$.}

. \label{phis}
\end{figure}

Here we will briefly review the solution for
$\phi$, and supplement it with some further details. The
LG-equation for $\phi(\rho)$, assuming that $\lambda=0$, is given
by,
\begin{equation}
\bigg[-\nabla_{\rho}^{2}-1+\gamma_{s}[f(l\rho)]^{2}+\phi^{2}\bigg]\phi=0,\label{scaledphi}
\end{equation}
 We shall now be interested in solving the linearized version of the
above equation, which is justified for $\gamma_{s}>1$. For $\rho\ll l^{-1}$,
this becomes equivalent to solving the problem
\begin{equation}
\bigg[-\nabla_{\rho}^{2}-1+\gamma_{s}\left(Cl\rho\right){}^{2}\bigg]\phi=0.\label{phismalldist}
\end{equation}
 This is identical to solving the Schr\"{o}dinger equation for the 2D
quantum harmonic oscillator. We know that $\phi'(\rho=0)=0$. The
solution for this equation is given by
\begin{equation}
\phi(\rho)=e^{-\sqrt{\gamma_{s}C^{2}l^{2}}\rho^{2}/2}\frac{{\cal {L}}_{a}(\sqrt{\gamma_{s}C^{2}l^{2}}\rho^{2})}{{\cal {L}}_{a}(0)},~~~a=\frac{1-2\sqrt{\gamma_{s}C^{2}l^{2}}}{4\sqrt{\gamma_{s}C^{2}l^{2}}},
\end{equation}
 where ${\cal {L}}_{n}(x)$ are the Laguerre polynomials. The profiles
of $\phi(\rho)$ for a few different values of $\gamma_{s}C^{2}l^{2}$
are shown in Fig.{\ref{phis}}.

%%%%%%%%%%%%%%%%%%%%%%%%%%%%%%%%%%%%%%%%%%%%%%%%%%%%%%%%%%%%

%%%%%%%%%%%%%%%%%%%%%%%%%%%%%%%%%%%%%%%%%%%%%%%%%%%%%%%%%%%%

At this point, we can also describe how we obtained the equation for
the phase boundary between regions I and II in the phase diagram (Fig.
\ref{lvsg}). In this case, $\phi$ is non-zero only very close to
the center of the core. We can therefore expand $\psi$ around $\rho=0$
and keep only the leading order term (Eq. \ref{asymp}). Eqn.{\ref{phismalldist}}
can be re-written as,
\begin{equation}
\bigg[-\frac{\nabla_{\rho}^{2}}{2}+\frac{\gamma_{s}C^{2}l^{2}}{2}\rho^{2}\bigg]\phi=(\frac{1}{2}+\epsilon)\phi.
\end{equation}
 Non-trivial solutions exist for $\epsilon\leq0$. The above equation
is the Schr\"{o}dinger equation for a quantum harmonic oscillator
in two dimensions with $m=1,\omega^{2}=\gamma_{s}(Cl)^{2}$. Then
the smallest eigenvalue which corresponds to the zero point energy
of the oscillator leads to the following equation for the curve
\begin{equation}
\gamma_{s}=\frac{1}{4(Cl)^{2}}
\end{equation}
 in the limit of small $l$. \\
 On the other hand, for $\rho\gg l^{-1}$, we have to solve
\begin{equation}
\bigg[-\nabla_{\rho}^{2}-1+\gamma_{s}\bigg(1-\frac{1}{(\rho l)^{2}}\bigg)\bigg]\phi=0,\label{philargedist}
\end{equation}
 from which we see that $\phi(\rho)\sim\rho^{-1/2}\exp(-\sqrt{\gamma_{s}-1}\rho)$
. However, there is a \textit{fine-tuned} point at $\gamma_{s}=1$,
at which the field $\phi$ is critical far away from the vortex core.
The full equation for $\phi$ becomes
\begin{equation}
\bigg[-\nabla_{\rho}^{2}-\frac{1}{(\rho l)^{2}}+\phi^{2}\bigg]\phi=0,\label{finephi}
\end{equation}
 %EBthe solution for which is given by which admits the solution
$\phi(\rho)=\sqrt{1+l^{2}}/(l\rho)$. %EBHowever, the above equation is a non-linear differential equation which admits multiple solutions.
Note that the $1/\rho$ solution can be obtained only for specific
boundary conditions. For generic boundary conditions, with $\phi\rightarrow0$
as $\rho\rightarrow\infty$, the solution is nevertheless asymptotic
to $\sqrt{1+l^{2}}/(l\rho)$ at large $\rho$. If $\phi(1)\ll1$,
for instance, then $\phi(\rho)\sim A\ln\left(\rho_{0}/\rho\right)$
at intermediate values of $\rho$, where $A$ and $\rho_{0}$ are
constants. $\phi(\rho)$ crosses over to $\phi(\rho)\sim1/\rho$ at
radii of the order of $\rho^{\star}\sim1/A$ (see Appendix \ref{appasym}).
This behavior reflects itself in the asymptotic decay of the anisotropy
of the field $\psi$ away from the vortex core, as we saw in Sec.
\ref{min}.

%EB, which requires a lot of fine tuning. . If we solve the above differential equation as an initial value problem with  and  specified for some , then unless , the solution blows up to either of . \\
%EBLet us now analyze this equation in the limit where  is small. It is then safe to ignore the  term %and for , the solution for  is then given by , where  is determined by the boundary condition at %small . It therefore seems that there exists a cross-over from this  solution to the power law %solution. We can estimate the value of this crossover distance, , by plugging back the logarithmic %solution back into Eqn. and approximating  as roughly a constant. This then gives us,  Therefore, if %we start from a regime where  is small, we expect to obtain a logarithmic decay over a long range of %the order of  which finally crosses over to a  form asymptotically. We'll dwell upon this point %again when we minimize the full LG free energy corresponding to this critical case.
% Therefore, at this point we see that the nematic order is quasi-long ranged, a fact that will have %some interesting consequences to be explored later. It is worth mentioning here that this case will %turn out to have qualitative differences from the case with localized nematic order when we include %the effect of anisotropy.

\subsection{Coexistence of superconductivity and nematic order}

\label{coex} In this subsection, we are interested in analyzing
region III of the phase diagram, in which superconductivity and
nematicity coexist even far away from the core. Let us assume, for
simplicity, that far away from the core $\phi$ can be replaced by
a constant. The effect of a constant $\phi$ is to render the
effective masses in the two directions different. Therefore, if we
re-scale the coordinates as
\begin{eqnarray}
x'=\frac{x}{\sqrt{1+\alpha}}\nonumber \\
y'=\frac{y}{\sqrt{1-\alpha}},
\end{eqnarray}
 where $\alpha=2\lambda\phi l^{2}$, then this problem now becomes
identical to the isotropic problem we had solved in the beginning
of section {\ref{mod}}. The solution for $\Psi_{0}$ can then be
written in terms of the new coordinates as,
\begin{eqnarray}
\Psi_{0} & = & \frac{f(r')}{r'}(x'+iy'),\nonumber \\
f(r')=c\bigg(1 & - & \frac{1}{2l^{2}c^{2}r'^{2}}\bigg),~~~~c=\sqrt{1-\frac{\gamma^{2}\phi^{2}}{\gamma_{s}}}
\end{eqnarray}
 Note that due to the presence of the background nematic order, $\Psi_{0}$
does not tend to $1$ asymptotically. We now go back to our original
coordinate system $x,y$ by expanding the above result to linear order
in $\alpha$. Then we get,
\begin{equation}
\psi=c\bigg(1-\frac{1}{2l^{2}c^{2}r^{2}}\bigg)e^{i\theta}-\frac{\alpha c}{4}\bigg(1+\frac{1}{2l^{2}c^{2}r^{2}}\bigg)e^{-i\theta}+\frac{\alpha c}{4}\bigg(1-\frac{3}{2l^{2}c^{2}r^{2}}\bigg)e^{3i\theta}\label{psireg3}
\end{equation}
 In the above expression, the first bracket corresponds to $\Psi_{0}$,
the second bracket corresponds to $\Psi_{-1}$ while the last one
represents $\Psi_{1}$. It is interesting to observe that asymptotically,
$\Psi_{1}$ and $-\Psi_{-1}$ approach the same constant value. We
observe this feature in Fig.(\ref{gam0p5l0p5}a). However, the harmonics
do not recover to their asymptotic value as a power law, which is a result of the boundary conditions
that were imposed while minimizing the free energy in the disk geometry
(see Appendix \ref{boundary}).

From Eqn. \ref{psireg3}, we can evaluate the form of $|\psi|^{2}$
and find that,
\begin{equation}
|\psi|^{2}=c^{2}\bigg(1-\frac{1}{l^{2}c^{2}r^{2}}\bigg)-\frac{\alpha}{l^{2}r^{2}}\cos(2\theta)+O(\alpha^{2})
\end{equation}
 Therefore, asymptotically, $|\psi|^{2}$ is isotropic and the anisotropy
decays as $\sim\alpha\cos(2\theta)/r^{2}$.

\subsection{Linearized GL analysis}

\label{lingl} In this section, we shall carry out an analysis of
the linearized LG equations, to give an analytical explanation for
some of the features that we have observed by carrying out the
full minimization. For the sake of simplicity, let us ignore the
feedback on $\phi$ resulting in the generation of the higher
harmonics and assume that $\phi$ is isotropic (i.e.
$\phi(\rho,\theta)=\phi(\rho)$). Then, the linearized LG equations
for the harmonics of $\psi$ can be written as,
\begin{eqnarray}
\frac{1}{l^{2}}\bigg(\partial_{\rho}^{2}+\frac{\partial_{\rho}}{\rho}-\frac{(2n+1)^{2}}{\rho^{2}}\bigg)\Psi_{n}(\rho) & + & \bigg(1-\frac{\gamma^{2}}{\gamma_{s}}\phi^{2}\bigg)\Psi_{n}(\rho)-2\Psi_{0}^{2}(\rho)\Psi_{n}(\rho)-\Psi_{0}^{2}(\rho)\Psi_{-n}(\rho)\nonumber \\
=-\lambda\phi(\rho)\bigg[\bigg(\partial_{\rho}^{2}-(4n-1)\frac{\partial_{\rho}}{\rho} & + & \frac{(4n^{2}-1)}{\rho^{2}}\bigg)\Psi_{n-1}(\rho)+\nonumber \\
\bigg(\partial_{\rho}^{2}+(4n+5)\frac{\partial_{\rho}}{\rho} & + & \frac{(2n+3)(2n+1)}{\rho^{2}}\bigg)\Psi_{n+1}(\rho)\bigg]\nonumber \\
+\lambda\partial_{\rho}\phi(\rho)\bigg[\bigg(\partial_{\rho}-\frac{2n-1}{\rho}\bigg)\Psi_{n-1}(\rho) & + & \bigg(\partial_{\rho}+\frac{2n+3}{\rho}\bigg)\Psi_{n+1}(\rho)\bigg]\label{harmonics_nd}
\end{eqnarray}
 There are some features of the problem that cannot be deduced from
a study of the linearized version of the problem, which include the
overall scale and sign of $\phi$ and the signs of the different
harmonics of $\psi$.

In the limit of $\rho\ll l^{-1}$, i.e. inside the vortex core, at
leading order $\Psi_{n}(\rho)\sim\rho^{a}$, where $a=|2n+1|$. This
is a necessary condition for the harmonics to be well behaved in the
limit of $\rho\to0$.

On the other hand, in the limit of $\rho\gg l^{-1}$, i.e deep inside
the superconducting region, the homogenous solution for the above
equation gives exponentially damped solutions for all the $\Psi_{n\neq0}$,
i.e. the anisotropy is short ranged. Moreover, the source term, which
is proportional to $\lambda\phi$ and is itself exponentially damped (Region II),
is also not strong enough to give rise to any long ranged solution.

However, when $\phi$ is critical (i.e. $\gamma_{s}=1$), the source
term leads to the presence of long tails in the harmonics. In the
regime where $\phi$ falls off logarithmically while $\Psi_{0}$ is
a constant, at leading order $\Psi_{\pm1}$ just follow $\phi$, i.e.
they also fall off logarithmically (with prefactors of equal magnitude
but opposite sign) and have a correction of the form $\ln\rho/\rho^{2}$.
On the other hand, when $\phi$ crosses over to the power law form,
at leading order $\Psi_{\pm1}$ also fall off as $\pm1/\rho$ with
a correction of order $1/\rho^{3}$.

\section{Conclusion}

We have studied the interplay between nematic order and superconductivity in the presence of a vortex.  If the nematic order coexists with superconductivity in the vicinity of a
vortex core, the coupling between the two order parameters leads
to an elongated shape of the core. We diskuss two distinct
scenarios: in one the nematic order coexists with
superconductivity everywhere (i.e., even far away from the vortex
core), whereas in the other the competition between the two order
parameter suppresses the nematic order in the bulk, and nematicity
only exists close to the core where the superconducting order
parameter is diminished. Both scenarios lead to an anisotropic
core. However, we show that they can, in principle, be
distinguished by the way the anisotropy of the superconducting gap
decays away from the core. If the nematicity exists only near the
core, the anisotropy in the superconducting gap decays
exponentially; if it exists throughout the sample, we expect the
gap anisotropy to decay as $1/r^2$, where $r$ is the distance from
the core. Moreover, there are qualitative differences in the shape
of the core in the two cases. In the former case, in which only
the core region is nematic, the contours of equal gap tend to be
more or less elliptical.  In the latter case, the contours of
equal gap tend to develop non-elliptical shapes with a four-petal
pattern. Therefore, analyzing the gap profiles
measured by STM around a vortex could reveal the nature of the
nematic ordering - whether it is localized at the vortex core, or
coexists with superconductivity in the bulk.

%If the spectrum resembles an ellipse and the anisotropy decays
%within the vortex core, then the nematicity is most likely
%localized within the core. On the other hand, if the spectrum
%develops a non-elliptical feature with the anisotropy decaying as
%a power law decay as a function of distance, then it is likely to
%be due to a background nematic order that exists throughout the
%sample.}

% {*}{*}{*}More to be added here after the full picture emerges{*}{*}{*}
So far, we have diskussed the structure of an isolated vortex at
the mean-field level. However, if the nematic ordering is favored
only within a vortex core, an isolated vortex cannot have static
nematic order, since either thermal or quantum fluctuations would
destroy such order. Static nematic order is only possible when the
density of vortices is finite. The coupling between the nematic
halos of different vortices scales as $J_\mathrm{eff} \sim
\exp[-d/(\sqrt{1-\gamma_s}l_\phi)]$, where $d\sim 1/\sqrt{B}$ is
the inter-vortex distance ($B$ is the applied magnetic field). The
system can be described by an effective two-dimensional transverse
field Ising model with a spin-spin interaction $J_\mathrm{eff}$
and a $B$-independent transverse field. (Note that, unlike Ref.
\cite{Kivelson02}, we are considering a thin film, rather than a
three-dimensional system.) This model has a nematic transition at
a certain critical $B$, which should be seen, e.g., by measuring
the anisotropy of the vortex cores as a function of $B$. If an
external rotational symmetry breaking field exists, as is
presumably the case in FeSe due to the small orthorhombic lattice
distortion\cite{mcqueen}, the electronic nematic transition is
smoothed out. However, one still expects a sharp crossover as a
function of magnetic field if the orthorhombic distortion is
sufficiently weak.

The microscopic origin of the anisotropic vortex cores observed in
FeSe\cite{song11} remains to be understood. It is likely that it
originates from electronic nematicity rather than from the lattice
distortion, since the experimentally reported orthorhombic
distortion seems too small to produce such a large effect. The
electronic nematic order could have an orbital
character\cite{rajiv, kruger, ccchen2010, ming}. Alternatively, it
could arise from a field-induced magnetic ordering\cite{lake} at a
wavevector $(\pi,0)$ or $(0,\pi)$ in the one iron unit cell, which
is necessarily accompanied by a nematic component (similar to the
ordering in the iron arsenides). Although static ordering of this
type has not been observed in the iron selenides\cite{cava}, it remains to be
seen if they develop a static ordering in the presence of an applied magnetic
field. Neutron
scattering experiments revealed a magnetic resonance at this
wavevector in the superconducting state of
FeTeSe\cite{neutronFeSe}. Moreover, ordering at such wavevectors
nearly nests the electron and hole pockets, and therefore it is
expected to couple strongly to superconductivity, explaining why
the resulting anisotropy of the vortex cores is so large.

{\it Note added:} After this work was submitted for
publication, another manuscript\cite{congjun} that studied the
experimental features observed in FeSe \cite{song11} came to our
attention. In this paper, the authors study the effect of orbital
ordering on the vortex structure in a two band model, by solving
the Bogoliubov-de Gennes equations. This study is complementary to
our phenomenological Ginzburg-Landau approach.
\appendix
%dummy comment inserted by tex2lyx to ensure that this paragraph is not empty
%dummy comment inserted by tex2lyx to ensure that this paragraph is not empty
%dummy comment inserted by tex2lyx to ensure that this paragraph is not empty

\acknowledgements

This research was supported by the National Science Foundation
under grants DMR-1103860, DMR-0705472 and by a MURI grant from
AFOSR. D.C. thanks Gilad Ben-Shach for a critical reading of the manuscript and for his comments. D.C. also thanks the Physics department at Harvard University for an E.M. Purcell fellowship during 2010-11.

\section{Instabilities of the free energy}

\label{appins} An interesting feature associated with the LG
functional introduced in section \ref{mod} is that there is an
instability to a state with modulated $\psi$. This arises due to a
competition between two terms in the free energy, namely the
$\phi$ and $\phi^{2}$ terms. Let us suppose that $\phi$ does not
vary spatially and $\psi=\beta e^{iqx}$. Then at leading order,
the contribution to the free energy from $\phi$ is of the form
\begin{equation}
{\cal {F}}_{\phi}=\bigg(\frac{\gamma^{2}\beta^{2}}{2\gamma_{s}}-\frac{\gamma^{2}}{2\gamma_{s}^{2}}\bigg)\phi^{2}+\lambda q^{2}\beta^{2}\phi.
\end{equation}
 From the above expression, we see that for a sufficiently large $\lambda q^{2}$,
it becomes energetically favorable to gain energy from the second
term by condensing a large negative value of $\phi$. By
extremizing the above with respect to $\phi$, we obtain
$\phi_{m}=-\lambda\beta^{2}q^{2}/\bigg(\frac{\gamma^{2}\beta^{2}}{\gamma_{s}}-\frac{\gamma^{2}}{\gamma_{s}^{2}}\bigg)$.
Hence, the contribution to free energy from $\phi_{m}$ is
$\propto-\beta^2 \lambda^2 q^{4}$. This energy gain from a
non-zero $q$ always dominates over the energy cost of order $q^2$
for a sufficiently large $q$. In order to prevent this
instability, we have to add a term of the form
$\zeta|\nabla^{2}\psi|^{2}/2l^{4}$ to the free energy, which is
an allowed term from the underlying symmetry of the problem. We
now want to obtain
some restrictions on $\zeta$. \\
 First of all, $\zeta$ should be such that it prevents the instability.
This gives us a lower bound on the value of $\zeta$. At the same
time, $\zeta$ should be small enough so that it should not change
the physics significantly. This gives us an upper bound on the
value of $\zeta$. Therefore, we obtain,
\begin{equation}
\frac{\lambda^{2}\gamma_{s}^{2}l^{4}}{\gamma^{2}(\gamma_{s}-1)}<\zeta\ll1\label{ra}
\end{equation}
 The above expression is not valid when $\phi$ becomes critical,
i.e. when $\gamma_{s}=1,\beta=1$. In this case, we have to compare
$\phi$ with $\phi^{4}$. \\
 However, when we minimized the free energy in section \ref{min},
we did not have to include the above term with a finite $\zeta$ as for a sufficiently
small $\lambda$, the cutoff in $q$ arising from the discrete lattice
prevented this instability from showing up.

\section{Effect of boundary terms}

\label{boundary} In general, when we derive the GL equations from
the free energy, there is a surface term arising from the gradient
terms in the energy which can be ignored in the limit of an infinite
system size. However, for a finite sized system, the boundary term
does play an important role. Let us consider only the contribution
of the gradient term of the superconducting order parameter in the
free energy, in the absence of any nematic order. Then we have,
\begin{eqnarray}
F_{grad}=\int d^{2}r|\nabla\psi|^{2},\\
F_{tot}=F_{\mathrm{grad}}+F_{\mathrm{local}},
\end{eqnarray}
 where $F_{local}$ contains the usual $|\psi|^{2},|\psi|^{4}$ terms.
On varying $\psi^{*}$ by $\delta\psi^{*}$ in $F_{tot}$, we obtain
for a finite system (up to other variations due to $F_{local}$
denoted by $...$),
\begin{equation}
-\int
d^{2}r\delta\psi^{*}\nabla^{2}\psi+\int_{\mathrm{surface}}\delta\psi^{*}(\nabla\psi)\cdot\hat{n}ds+...=0,
\end{equation}
 where $\hat{n}ds$ is the area element, normal to the boundary. When
we solve for $\psi$ in the interior of the region, only the first
term contributes and the boundary term can be ignored. However,
when we solve for $\psi$ on the boundary, only the surface term
plays a role, since it can be thought of as appearing with an
infinite weight of the form $\int dr\delta(r-R)$ where $R$ is the
radius of the disk on which we are minimizing the free energy and
$\delta(...)$ is the Dirac-delta function. Therefore, in order to
solve for $\psi$, we have to solve for $-\nabla^{2}\psi+...=0$ in
the interior of the region subject to the boundary condition
$\nabla\psi\cdot \hat{n}|_{r=R}=0$ (Neumann boundary conditions).
\\
 Now in the presence of a constant nematic background ($\phi_{0}$),
the gradient term in the free energy is,
\begin{equation}
F_{grad}=\int dxdy\bigg[(1+\alpha)|\partial_{x}\psi|^{2}+(1-\alpha)|\partial_{y}\psi|^{2}\bigg],
\end{equation}
 where $\alpha=2\lambda\phi_{0}l^{2}$. As we did earlier, on carrying out the variation over $\psi^*$ this amounts
to solving for
$-(1+\alpha)\partial_{x}^{2}\psi-(1-\alpha)\partial_{y}^{2}\psi+...=0$
subject to the boundary condition, $\tilde{D}\psi\cdot \hat{n}=0$,
where
\begin{equation}
\tilde{D}\psi=\bigg((1+\alpha)\partial_{x}\psi,(1-\alpha)\partial_{y}\psi\bigg),~~~\hat{n}=(\cos\theta,\sin\theta).\label{bc}
\end{equation}
 In polar coordinates, this condition can be written as,
\begin{equation}
\partial_{r}\psi+\alpha\bigg[\cos2\theta\partial_{r}\psi-\frac{\sin2\theta}{r}\partial_{\theta}\psi\bigg]=0.
\end{equation}

These boundary conditions mean, in particular, that the current
perpendicular to the boundary is zero. In our numerical
calculations, we have used a disk geometry; therefore the
boundaries are found to have a significant effect whenever we are
considering a non-circularly symmetric solution, in particular in
the regime where the nematic order is non-zero even far away from
the core. We circumvent this problem, however, by taking a
sufficiently large system and considering the solution only close
to the vortex core, where the boundary effects are small.

\section{Asymptotics of $\phi$ in the critical case}
\label{appasym}
%%%%%%%%%%%%%%%%%%%%%%%%%%%%%%%%%%%%%%%%%%%%%%%%%%%%%%%%%%
\begin{figure}
\psfig{figure=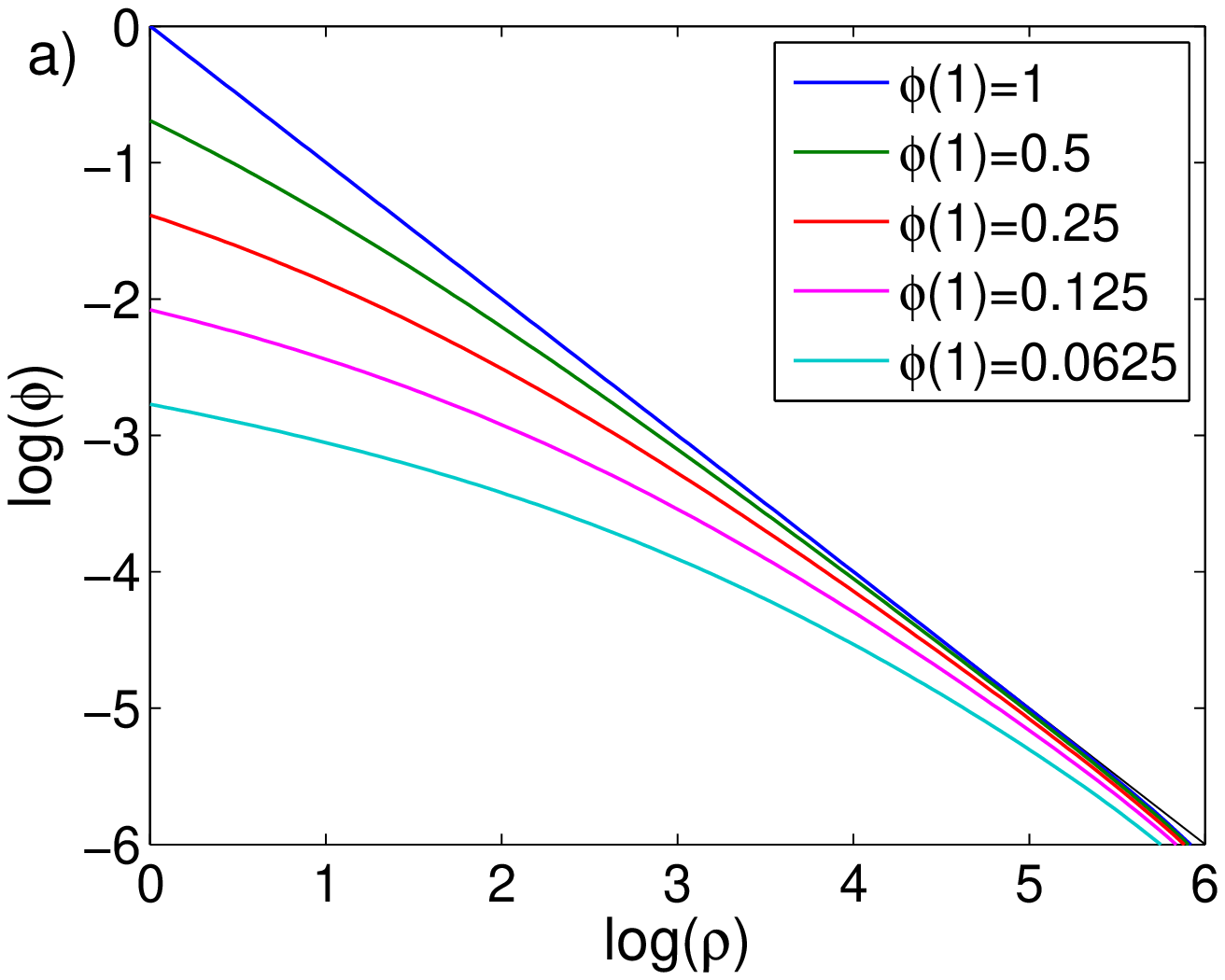,width=95mm} %\vspace{1mm}
\psfig{figure=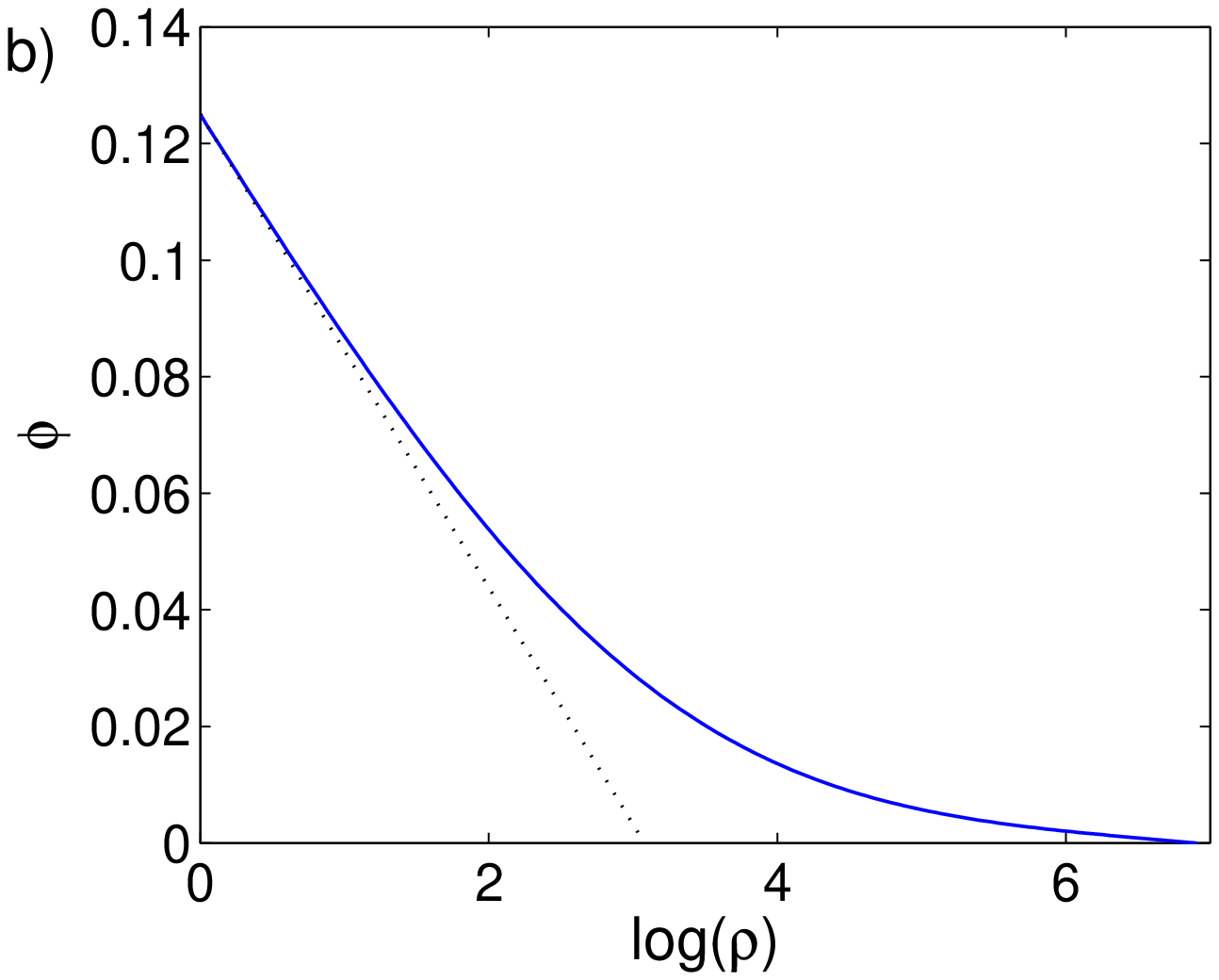,width=95mm} \vspace*{1mm}
 \caption{Numerical solutions of Eq. \ref{eq:phi_critical}. a) Solutions with boundary conditions
 $\phi(1000)=0$ and with various values of $\phi(1)$, on a log-log scale. b) Solution with boundary condition
 $\phi(1)=0.125$, $\phi(1000)=0$, on a semilog scale. The dashed line is a fit to the form $A+B\log(\rho)$ for
 small $\rho$.}
\label{fig:phi_r}
\end{figure}
%%%%%%%%%%%%%%%%%%%%%%%%%%%%%%%%%%%%%%%%%%%%%%%%%%%%%%%%%%%%%

In this appendix, we analyze the asymptotics of the field $\phi$
far away from the vortex core in the case $\gamma_{s}=1$, in which
the nematic order is critical. In this case, and for $\rho\gg1$,
the Landau-Ginzburg equation for $\phi$ (Eq. \ref{scaledphi}) becomes
\begin{equation}
\bigg[-\nabla_{\rho}^{2}+\phi^{2}\bigg]\phi=0.\label{eq:phi_critical}
\end{equation}
This non-linear equation admits the solution
$\phi\left(\rho\right)=1/\rho$ \cite{Kivelson02}. This solution is
valid, however, for specific initial conditions, e.g.,
$\phi(1)=1$, $\phi'(1)=-1$. Physically, the initial conditions for
Eq. \ref{eq:phi_critical} are determined by the details of the
vortex profile at short distances, determined by Eq.
\ref{scaledphi}. Nevertheless, one can make some general
statements about the asymptotic behavior of the solution. If, for
some arbitrary $\rho_{0}$ such that $\rho_{0}\gg 1/l$ (far from
the core), $\phi$ satisfies $\phi\left(\rho_{0}\right)\ll
1/\rho_{0}$, then it is justified to neglect the $\phi^{2}$ term
in Eq. \ref{eq:phi_critical}. Then, the solution close to
$\rho_{0}$ behaves as $\phi\left(\rho\right)\approx
A-B\ln\left(\rho/\rho_{0}\right)$, where $A$, $B$ are determined
by the initial conditions. This can only be valid, however, up to
a point $\rho_{*}$ at which
$\phi\left(\rho_{*}\right)\approx1/\rho_{*}$, i.e., at distances
which are much smaller than the length scale set by the initial
condition of Eq. \ref{eq:phi_critical}. At longer distances, we
expect a crossover to $\phi\left(\rho\right)\approx1/\rho$.

In Fig. \ref{fig:phi_r}, we present a numerical solution of Eq.
\ref{eq:phi_critical} with boundary conditions $\phi(1000)=0$ and
various values for $\phi(1)$. When $\phi(1)=1$, we get
$\phi(\rho)\approx 1/\rho$ (where the deviations are due to the
boundary condition at $\rho=1000$). For smaller $\phi(1)$, there
is an intermediate region where $\phi$ does not follow a power
law, eventually crossing over to $1/\rho$ at larger $\rho$.
$\phi(\rho)$ is approximately logarithmic in the intermediate
region, as shown in Fig. \ref{fig:phi_r}b.

Physically, we expect that $\phi<1$ (since $\phi=1$ corresponds to
the equilibrium value of $\phi$ in the absence of
superconductivity). Therefore, there is an intermediate
logarithmic region, which becomes parametrically large in the
limit of small $\phi$.

\end{document}